\documentclass[prc,preprint,showpacs,showkeys,nofootinbib,tightenlines]{revtex4}

\usepackage{graphicx}  %
\usepackage{bm}  %

\newcommand{\beq}{\begin{equation}}
\newcommand{\eeq}{\end{equation}}
\newcommand{\bea}{\begin{eqnarray}}
\newcommand{\eea}{\end{eqnarray}}
\newcommand{\ben}{\begin{eqnarray*}}
\newcommand{\een}{\end{eqnarray*}}

\renewcommand{\vec}[1]{{\mathbf #1}}

\newcommand{\nab}{\overrightarrow{\nabla}}

\newcommand{\Mlo}{M_{lo}}
\newcommand{\Mhi}{M_{hi}}

\begin{document}

\title{Narrow Resonances in Effective Field Theory}

\author{P.F. Bedaque}
\affiliation{Nuclear Science Division, 
             Lawrence Berkeley National Laboratory, Berkeley, CA 94720, USA}
\author{H.-W. Hammer}
\affiliation{Helmholtz-Institut f{\"u}r Strahlen- und Kernphysik
             (Abt.~Theorie), Universit{\"a}t Bonn, 53115 Bonn, Germany}
\author{U. van Kolck}
\affiliation{Department of Physics, University of Arizona,
          Tucson, AZ\ 85721, USA}
\affiliation{RIKEN-BNL Research Center, Brookhaven National Laboratory,
           Upton, NY\ 11973, USA}

\date{April 2, 2003}

\begin{abstract}
We discuss the power counting for effective field theories
with narrow resonances near a two-body threshold.
Close to threshold, the effective field theory is perturbative
and only one combination of coupling constants is fine-tuned.
In the vicinity of the resonance, a second, ``kinematic'' 
fine-tuning requires a nonperturbative resummation.
We illustrate our results in the case of 
nucleon-alpha scattering.
\end{abstract}

\smallskip
\pacs{21.45.+v, 25.40.Dn}
\keywords{Effective field theory, narrow resonances}
\maketitle

The last ten years have seen the development
of effective field theories (EFTs) for systems of few nucleons
\cite{ARNPSreview,NN99}.
Nucleons in light nuclei have typical momenta that are small
compared to the characteristic QCD scale of 1 GeV.
At these low momenta, QCD can conveniently be
represented by a hadronic theory containing
all possible interactions consistent with the QCD symmetries.
It is crucial to formulate a power counting that
justifies a systematic and controlled truncation of
the Lagrangian according to the desired accuracy.
Nuclei offer a non-trivial challenge because 
one wants such a perturbative expansion
in addition to the non-perturbative treatment of certain
leading operators, which is required by the existence of 
shallow bound states. 
By now, two-, three- and four-nucleon systems have been studied
with EFT.
While much remains to be understood, many successes have been
achieved \cite{ARNPSreview,NN99}.

The extension of EFTs to larger nuclei faces computational
challenges, as do other approaches.
As a first step \cite{halos} in this extension, we can specialize
to very low energies where clusters of nucleons 
behave coherently. Even though many interesting issues of 
nuclear structure are avoided, we can still describe anomalously 
shallow (``halo'') nuclei and some reactions of astrophysical interest.

Reactions involving more complex nuclei are frequently characterized
by narrow resonances near threshold.
One example, the $p_{3/2}$ resonance in neutron-alpha ($n\alpha$)
scattering, was considered in Ref.~\cite{halos}.
A good description of the data throughout the resonance
region was found
at the expense of the resummation of two operators
in addition to the unitarity cut.
Such a resummation requires {\it two} fine-tunings,
which is somewhat surprising.
Here we generalize the analysis of shallow, narrow resonances
in EFT. We discuss the power counting for resonances
in any partial wave, and clarify
the scope of unitarization. 
We again use $n\alpha$ scattering as an example,
in order to make the comparison with Ref.~\cite{halos} explicit.

We consider a two-body scattering with reduced mass $\mu$
and energy $E=k^2/2\mu$ in the center-of-mass frame.
Resonance behavior arises 
when the $S$-matrix has a pair of poles in the two lower
quadrants of the complex $k$ plane.
The projection of $S$ in the partial wave $l$ where
the resonance lies can be written
\begin{eqnarray}
S_l &=& -\frac{k+k_+}{k-k_+}\, \frac{k+k_-}{k-k_-} \, s_l(k) \nonumber \\
    &=& -\frac{E-E_0-i\Gamma(E)/2}{E-E_0+i\Gamma(E)/2} \, s_l(k).
\label{smat}
\end{eqnarray}
Here $k_\pm= \pm k_R +i k_I$ with $k_I<0$ are the pole positions,
$s_l(k)$ is a smooth function in the energy region
under consideration,
$E_0 =(k_R^2 + k_I^2)/2\mu$ is
the position of the resonance ---defined as the energy
where the corresponding phase shift crosses $\pi/2$---
and $\Gamma(E)/2= - k k_I/\mu$ is referred to as the half-width of the 
resonance.
A narrow resonance is one for which $\Gamma(E_0)/2E_0\ll 1$,
that is, for which the poles are near the real axis,
$|k_I/k_R|\ll 1$. 

We say the resonance is shallow
if $|k_\pm| \equiv M_{lo} \ll M_{hi}$, 
where $M_{hi}$ is the characteristic scale of the underlying theory,
such as the energy scale of excitations of the particles under consideration
and the mass of exchanged particles.
A physical example of a shallow, narrow resonance is 
the $p_{3/2}$ resonance in $n\alpha$
scattering, which in our fit \cite{halos}
has $\Gamma(E_0)/2\simeq 0.3 \, {\rm MeV} \ll E_0 \simeq 0.8 \, {\rm MeV}
\ll E_\alpha \simeq 20\, {\rm MeV}$,
or $k_I \simeq -6 \, {\rm MeV} \ll k_R \simeq 34 \, {\rm MeV}
\ll \sqrt{m_N E_\alpha} \simeq 140 \, {\rm MeV}$,
where $E_\alpha$ is the excitation energy of the $\alpha$ core
and $m_N$ is the nucleon mass.
(Similar values can be found in other fits \cite{tunl}.)
Another example of a shallow resonance
can be found 
in the $s$-wave channel of a two-range Gaussian
potential (see, {\it e.g.}, Ref.~\cite{polemodels}).

Physics at a momentum scale $Q\sim M_{lo}$ can be described by an
EFT containing as degrees of freedom only the scattering particles.
For notational simplicity we take the two particles to be identical, 
with mass $m=2\mu$ and no spin.
Generalization to other situations is straightforward.
In an EFT, observables are independent of the choice of fields.
It proves convenient to introduce a field $d$ ---the ``dimeron''---
with the quantum numbers of the resonance \cite{halos},
as can be done for a shallow $s$-wave bound state  \cite{Kap97}.

Let us first consider the case where the resonance is in the $l=1$ state.
The auxiliary field $d_i$ has spin $1$ and the Lagrangian reads,
\begin{equation}
\label{lagdPwave}
{\cal L}=\psi^\dagger \biggl[i\partial_0 + \frac{\nab^{\,2}}{2m}\bigg]\psi
         + \eta_1 d^\dagger_{i}
           \bigg( i\partial_0+\frac{\nab^{\,2}}{4m}-\Delta_1 \bigg)
           d_{i} 
        + \frac{g_1}{4}
           \left(d^\dagger_{i}
           \psi \tensor\partial_{i} \psi 
         +{\rm H.c.}\right)+\ldots\,,
\end{equation}
where the sign $\eta_1=\pm 1$, the parameters $g_1\equiv\sqrt{4\pi \alpha_1}$ 
and $\Delta_1$ 
are fixed from matching with the underlying theory or directly from the data,
$\tensor\partial_{i}$ is a shorthand notation for
$\roarrow\partial_{i} -\loarrow \partial_{i}$,
and the dots represent terms with more derivatives.
An EFT without the dimeron field can be obtained
by performing the Gaussian path integral over $d_i$.

\begin{figure}[tb]
\begin{center}
\includegraphics[width=5in,angle=0,clip=true]{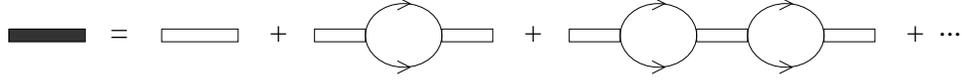}
\end{center}
\vspace*{-0pt} 
\caption{The full dimeron propagator (thick shaded line) is
obtained by dressing the bare dimeron propagator (double solid
line) with particle bubbles (solid lines) to all orders.}
\label{fig:auxprop}
\end{figure}

Introducing
${\tilde k}^2=mp_0-\vec{p}^{2}/4$,
the bare propagator for a dimeron of energy $p_0$ and momentum $\vec{p}$ 
is given by
\beq
\label{dbarel1}
iD_{l}^{(0)}(p_0,\vec{p})_{ij}=
\frac{i\eta_1 m}{{\tilde k}^2-m\Delta_1 +i\epsilon}\,\delta_{ij}.
\eeq
The full dimeron includes bubbles 
---shown diagrammatically in Fig.~\ref{fig:auxprop}---
generated by $d^\dagger \psi\psi$ interactions,
as well as insertions stemming from terms with more derivatives. 
The two-particle $T$-matrix can be obtained from the full dimeron 
propagator by attaching external particle legs. 

Before discussing the power counting, we compute the two-particle 
scattering amplitude including 
only the interactions explicitly shown in Eq.~(\ref{lagdPwave}). 
We work in the center-of-mass frame of the two particles, where
we denote the relative incoming (outgoing)
momenta by $\vec{k}\;(\vec{k'})$ and  the scattering angle by $\theta$.
The total energy is simply $k^2/m$, and 
${\tilde k}^2=k^2$. 
The result for the scattering amplitude is
\begin{equation}\label{TPwave}
T(k,\cos\theta) =\frac{12\pi}{m}k^2\cos\theta \left[
            \eta_1 \frac{12\pi\Delta_1}{mg_1^2}-\frac{2L_3}{\pi}
            -k^2\left(\eta_1 \frac{12\pi}{m^2g_1^2}+\frac{2L_1}{\pi}\right)-
            ik^3\right]^{-1}\,,
\end{equation} where 
\begin{equation}
L_{n}=\int dq \, q^{n-1}
\end{equation}
is an integral proportional to $\Lambda^{n}$, with
$\Lambda$ an ultraviolet momentum cutoff.
Matching Eq.~(\ref{TPwave}) to the effective-range expansion for
the scattering amplitude,
\begin{equation}\label{TeffectiverangePwave}
T(k,\cos\theta)=\frac{12\pi}{m}k^2\cos\theta\left[
            -\frac{1}{a_1}
            +\frac{r_1}{2}k^2+\ldots
            -ik^3\right]^{-1}\,,
\end{equation}\noindent 
the parameters $\Delta_1$ and $g_1$ in the Lagrangian (\ref{lagdPwave})
can be determined in terms of the effective-range parameters, for instance 
\begin{eqnarray}\label{matchingPwave}
	\frac{1}{a_1}&=& \frac{2L_3}{\pi}-\eta_1\frac{12\pi\Delta_1}{mg_1^2}
\,,\nonumber\\
	-\frac{r_1}{2}&=&\eta_1\frac{12\pi}{m^2g_1^2}+\frac{2L_1}{\pi}\,.
\end{eqnarray}

Dimensional analysis suggests that the typical size (in order of magnitude) 
for the effective-range parameters $a_1, r_1, \ldots$ is given by the 
appropriate power of the momentum scale $\Mhi$ where the effective theory 
breaks down. For instance, if the interaction between the particles is 
described by a potential of depth $\sim M_{hi}$ and range $\sim 1/\Mhi$, 
one would expect $a_1\sim 1/\Mhi^3$
and $r_1\sim \Mhi$. In particular, a resonance or bound state, if present, 
generally occurs at the momentum scale $\Mhi$. These cases have already 
been considered in Ref.~\cite{halos} and we have nothing to add to it here. In 
some systems, however, the interactions are finely tuned in such a way as to 
produce a resonance close to threshold, at a scale $\Mlo$ much smaller 
than $\Mhi$, violating the naive dimensional-analysis estimate. This situation 
can occur when one or more of the effective-range parameters have unnatural 
sizes related to the low-momentum scale $\Mlo$.
In Ref.~\cite{halos}, the situation when $a_1\sim 1/\Mlo^3$
and $r_1\sim \Mlo$ was 
analyzed. Note that this requires that {\it two} combinations of constants, 
$\Delta_1/g_1^2$ and $1/g_1^2$, be fine-tuned against the large values of 
$L_n\sim (\Mhi)^n$ in order to produce a result containing powers of the small
scale $\Mlo$ (see Eq.~(\ref{matchingPwave})). 
Assuming this scaling for the effective-range parameters, we see that all 
three terms of the effective-range expansion in 
Eq.~(\ref{TeffectiverangePwave}) are of the same order
for momenta $k\sim \Mlo$ and are retained in the leading-order 
expansion of the effective theory. Assuming that no further fine-tuning 
occurs and the higher effective-range parameters have their natural size 
determined by the high-momentum scale $\Mhi$, these terms are suppressed 
by powers of $\Mlo/\Mhi$ and are subleading.

In this paper we suggest a different scaling, in which $a_1\sim 1/(\Mlo^2
\Mhi)$, $r_1\sim \Mhi$ (and the other effective-range parameters scale 
with the appropriate power of $\Mhi$). This requires only {\it one} 
combination of constants, namely $\Delta_1/g_1^2$, to be fine-tuned. 
Systems obeying this scaling are not generic. However, they are more
likely to occur than the ones with the scaling assumed
in Ref.~\cite{halos}, since only one accidental fine-tuning is required. 
If the underlying theory cannot be
solved, the appropriate scaling for a specific physical system 
can be determined from the data, that is,
from the numerical values of the effective-range 
parameters. However, such a phenomenological determination is not always 
unique and/or different scalings might apply in different kinematic regions.

With the new scaling proposed above, the first two terms in the 
square brackets in Eq.~(\ref{TeffectiverangePwave}) are of the same order
for momenta $k\sim\Mlo$.
The term stemming from the unitarity cut, $ik^3$, is suppressed by one power of 
$\Mlo/\Mhi$ and is, therefore, subleading. 
The remaining terms in the effective-range expansion are even more suppressed.
A low-energy expansion in powers of $k/\Mhi$ that sums up all terms of order 
$k/\Mlo$ can then be obtained by taking the propagator in Eq.~(\ref{dbarel1}) 
as the leading-order term and the effects of loops and higher-derivative 
interactions as higher-order corrections.

At leading order the difference between the two scalings is the presence  
of the unitarity-cut term $\sim ik^3$. This difference disappears if 
instead of considering generic momenta $k$ of order $\Mlo$ we focus onto a 
narrow region around the position of the resonance at $k=\sqrt{2/a_1r_1}$. 
Due to the near cancellation between the two leading terms within a window of 
size  $\Delta k = 2/a_1r_1^2$
around the pole, the unitarity-cut term has 
to be resummed to all orders, and provides a width to the resonance. 
In this 
kinematic range there are two fine-tunings: one implicit in the 
short-distance physics leading to the unnatural value of $a_1$, 
and another one 
explicitly caused by the choice of kinematics close to the position of the 
pole. The size of the region where this resummation is necessary is of the 
order of $\Mlo^2/\Mhi$,
so, unless the scales
$\Mlo$ and $\Mhi$ are very separated, it may constitute a numerically 
significant part of the region of validity of the effective field theory 
($k\ll\Mhi$).

The extension of the situation described above to the case of resonances 
in higher partial waves is straightforward. 
The auxiliary field will carry $l$ vector indices
and we will use a short-hand notation to denote this set of indices:
 $d_{\{i\}}=d_{i_1 \cdots i_l}$ has $l$ integer indices $i_1, \ldots, i_l$.
Analogously, we use 
$\partial^l_{\{i\}}$ for the angular momentum $l$ 
part of $\partial_{i_1} \cdots \partial_{i_l}$,\footnote{ 
In the $l=2$ case, for instance, $\partial^2_{\{i\}}\propto (\partial_i
\partial_j-\partial^2\delta_{ij}/3$).} 
and 
$\delta_{\{i\} \{j\}}$ for the angular momentum $l$ part of 
$ \delta_{i_1 j_1} \cdots \delta_{i_l j_l}$.

The corresponding Lagrangian is
\bea
\label{lagd}
{\cal L}&=&\psi^\dagger \biggl[i\partial_0 + \frac{\nab^{\,2}}{2m}\bigg]\psi
         - \eta_l \Delta_l d^\dagger_{\{i\}} d_{\{i\}}
         + \sum_{n=1}^l r_n
           d^\dagger_{\{i\}}
           \bigg( i\partial_0+\frac{\nab^{\,2}}{4m}\bigg)^n
           d_{\{i\}}
\nonumber\\
        &&+ \frac{g_l}{4}
           \left(d^\dagger_{\{i\}}
           (\psi \tensor\partial^l_{\{i\}} \psi) 
         +{\rm H.c.}\right)+\ldots
\eea
Here $\eta_l$, $g_l\equiv\sqrt{4\pi \alpha_l}$ 
and $\Delta_l$ are the generalizations of the 
parameters $\eta_1$, $g_1$ 
and $\Delta_1$ in Eq. (\ref{lagdPwave}).
We also show explicitly dimeron
kinetic terms. The first ($n=1$) is simply the term displayed earlier, 
with $r_1 = \eta_l$.
The others ($n\ge 2$) can, 
alternatively, be eliminated by a $d$-field redefinition 
in favor of $d^\dagger \psi\psi$ 
interactions with derivatives.
As before, the EFT without the dimeron field can be obtained
by performing the Gaussian path integral over $d_{\{i\}}$.

The bare dimeron propagator, 
\beq
\label{dbare}
iD_{l}^{(0)}(p_0,\vec{p})_{\{i\} \{j\}}=
\frac{i\eta_l m}{{\tilde k}^2-m\Delta_l +i\epsilon}\,\delta_{\{i\} \{j\}},
\eeq
can generate two real poles,
which will be shallow provided
\begin{equation}
\Delta_l \sim \frac{M_{lo}^2}{m}. 
\label{pc1}
\end{equation}
The bubbles introduce unitarity corrections which
can dislocate the poles to the lower half-plane.
The resonance
will be narrow if the EFT is perturbative
in the coupling $\alpha_l$.
This will be so if 
\begin{equation}
m^2 \alpha_l 
\sim \left\{\begin{array}{ll}
     \frac{M_{lo}^{2}}{M_{hi}}
     &   l=0, \\ \\
     \frac{1}{M_{hi}^{2l-1}},
     &   l> 0
            \end{array} \right.
\label{pc2}
\end{equation}
(or weaker).
In this case, a loop is suppressed
by $M_{lo}/M_{hi}$ for $l=0$ and 
by $(M_{lo}/M_{hi})^{2l-1}$ for $l> 0$.
This is because adding a bubble means extra factors of 
$g_l^2 Q^{2l}$ from the two vertices,
$Q^3/4\pi$ from the integration,
$m/Q^2$ from the two-particle state,
and $m/Q^2$ from the dimeron propagator.
Waves that have no poles near threshold have all EFT parameters
scaling with $M_{hi}$ according to their mass dimensions.

For systems whose parameters scale in this way,
the dominant contribution to the two-body $T$-matrix
comes from the bare dimeron propagator (\ref{dbare}).
In the center-of-mass frame, the poles are at $k_\pm= \pm \sqrt{m \Delta_l}$.
If $\Delta_l>0$, then $k_I=0$ and $k_R \sim M_{lo}$, as desired.

The first corrections can be calculated from
the one-loop diagram in Fig. \ref{fig:auxprop}
and the corresponding counterterms.
The self-energy given by the particle bubble is
\bea
-i\Sigma_l (p_0,\vec{p})_{\{i\}\{j\}}&=& 
g_l^2 \int \frac{d^4 q}{(2\pi)^4}
\frac{\{q_{i_1} \cdots q_{i_l} q_{j_1} \cdots q_{j_l}\}}
{\left(\frac{p_0}{2}+q_0-\frac{(\vec{p}/2+\vec{q})^2}{2m}
+i\epsilon\right)\left(\frac{p_0}{2}-q_0-\frac{(\vec{p}/2-\vec{q})^2}{2m} 
+i\epsilon\right)}\nonumber \,,\\
&=& 
\frac{i m \alpha_l}{(2l+1)}
\left\{\frac{2}{\pi}
       \sum_{n=0}^{l} L_{2l-2n+1} {\tilde k}^{2n}
       +i {\tilde k}^{2l+1} \right\}
\delta_{\{i\} \{j\}}\,,
\label{sigma}
\eea
where $\{q_{i_1} \cdots q_{i_l} q_{j_1} \cdots q_{j_l}\}$ is the angular 
momentum $l$ part of $q_{i_1} \cdots q_{i_l} q_{j_1} \cdots q_{j_l}$.
The first correction to the dimeron propagator is then
\bea
\label{dfirst}
iD_{l}^{(1)}(p_0,\vec{p})_{\{i\} \{j\}} 
&=& iD_l^{(0)}(p_0,\vec{p})_{\{i\} \{k\}}
    (-i\Sigma_l(p_0,\vec{p})_{\{k\} \{l\}}) 
    iD_l^{(0)}(p_0,\vec{p})_{\{l\} \{j\}}
\nonumber\\
&=& i\left(\frac{im\eta_l}{{\tilde k}^{2} -m\Delta_l +i\epsilon}\right)^2
\left\{\eta_l \left(\Delta_l-\Delta_l^{R}\right)
       -\frac{\eta_l}{m} \left(1-\frac{\alpha_l}{\alpha_l^{R}}\right)
                           {\tilde k}^{2n}\right.
\nonumber\\
&&
\left.       +\sum_{n=2}^{l} \frac{r_n^{R}}{m^n} 
                      {\tilde k}^{2n}
       +\frac{i}{(2l+1)} m \alpha_l 
        {\tilde k}^{2l+1} \right\}
\,\delta_{\{i\} \{j\}}.
\eea
Here $\Delta_l^{R}$ is the finite part of 
\beq
\Delta_l(\Lambda)=
\eta_l \frac{1}{(2l+1)}\frac{2}{\pi} m \alpha_l(\Lambda) L_{2l+1}(\Lambda)
+\Delta_l^{R}\,,
\eeq
$\alpha_l^{R}$ is the finite part of
\beq
\alpha_l(\Lambda)=\alpha_l^{R}
\left( 1- \eta_l\frac{1}{(2l+1)}\frac{2}{\pi} m^2
 \alpha_l(\Lambda) L_{2l-1}(\Lambda) 
\right)\,,
\eeq
and the $r_n^{R}$ are the finite parts of the counterterms
\beq
r_n(\Lambda)=r_n^{R}-
\frac{1}{(2l+1)} \frac{2}{\pi} m^{n+1}\alpha_l(\Lambda) 
      L_{2l-2n+1}(\Lambda)\,.
\eeq

This procedure can be continued to higher orders in an obvious way.
As we have argued, each term is smaller than the previous one
by a power of $M_{lo}/M_{hi}$.
Nevertheless,
it is easy to see that the sum of the diagrams in  Fig. \ref{fig:auxprop}
forms a geometric series
\bea
iD_l(p_0,\vec{p})&=&iD_l^{(0)}(p_0,\vec{p})+iD_l^{(1)}(p_0,\vec{p})
+\ldots\nonumber\\
&=&iD_l^{(0)}(p_0,\vec{p})
\left(1-D_l^{(1)}(p_0,\vec{p})/D_l^{(0)}(p_0,\vec{p})\right)^{-1}\,,
\label{geo}
\eea
where the indices have been suppressed. 
The error induced by this resummation is of higher order.
So to subleading order we can write
\beq
iD_l(p_0,\vec{p})_{\{i\} \{j\}}=
i\eta_l m \delta_{\{i\} \{j\}}
\frac{\alpha_l^{R}}{\alpha_l}
\left\{-m\Delta_l^{R}+ {\tilde k}^2
+\eta_l m \sum_{n=2}^l \frac{r_n^{R}}{m^n} {\tilde k}^{2n}
+\frac{i}{(2l+1)} \eta_l m^2 \alpha_l^{R} 
{\tilde k}^{2l+1}\right\}^{-1}
\, .
\label{sum}
\eeq

In the center-of-mass frame, using the shorthand notation 
$\{k_{i_1}\ldots k_{i_l} k'_{j_1}\ldots k'_{j_l}\}$ defined above,
the $l$-wave projection of the $\psi \psi$ scattering amplitude is
\bea
T_l(\vec{k}',\vec{k})&=& -4\pi \alpha_l \{k_{i_1}\ldots k_{i_l} 
k'_{j_1}\ldots k'_{j_l}\}  D_l(k^2/m,\vec{0})_{\{i\} \{j\}}\nonumber\\
&=& 
\frac{4\pi}{ m} (2l+1) k^{2l} P_l(\cos\theta)
\left\{\frac{\eta_l (2l+1) \Delta_l^{R}}{m \alpha_l^{R}}-
 \frac{\eta_l (2l+1)}{m^2 \alpha_l^{R}}{k}^2 \right. \nonumber\\
 && \qquad \left.
 -\frac{\eta_l (2l+1)}{\alpha_l^{R}} \sum_{n=2}^l
   \frac{r_n^{R}}{m^{n+1}}{k}^{2n} -i {k}^{2l+1}\right\}^{-1}
\, ,
\label{fullt}
\eea
where $P_l(\cos\theta)$ is a Legendre polynomial.
From this the $S$-matrix (\ref{smat}) and all two-body observables follow.
Note that any reference to the cutoff has dropped:
only the renormalized quantities appear in  Eq. (\ref{fullt}).
From now on we drop the superscript $^{R}$.

The first two terms in the denominator of Eq. (\ref{fullt})
come from the bare propagator and give the two poles in the real axis,
\beq
k_\pm = \pm k_R^{(0)} \sim M_{lo},
\eeq
so that $E_0\sim M_{lo}^2/M_{hi}$ in Eq. (\ref{smat}).
They give rise to an $l$-wave scattering ``length''
\beq
a_l= -\eta_l \frac{m \alpha_l}{(2l+1) \Delta_l} 
\sim \left\{\begin{array}{ll}
            \frac{1}{(2l+1)} \frac{1}{M_{hi}},
           &  l=0, \\ \\
            \frac{1}{(2l+1)} \frac{1}{M_{hi}^{2l-1} M_{lo}^2},
           &  l> 0,
            \end{array} \right.
\eeq
and an $l$-wave effective ``range''
\beq
r_l= -\eta_l\frac{2\,(2l+1)}{m^2 \alpha_l} 
\sim \left\{\begin{array}{ll}
            (2l+1) \frac{M_{hi}}{M_{lo}^{2}},
           &  l=0,\\ \\
            (2l+1) M_{hi}^{2l-1},
           &  l> 0,
            \end{array} \right.
\eeq
The product $a_l r_l= 1/(m \Delta_l) \sim 1/M_{lo}^2$
is large in order to give a shallow pole, 
and indicates a fine-tuning.
However, only
one of the two quantities needs to be anomalously large.
For $l=0$ that is $r_0$.
(The other possibility, $a_0$, leads to a shallow virtual
or real bound state, and was discussed in Ref.~\cite{vKo99}.)
For $l\ge 1$
we are here considering systems where this quantity is $a_l$,
although the other possibility could be considered as well.

The next two terms in the denominator of Eq. (\ref{fullt})
are the first corrections.
The last is a consequence of unitarity. Being imaginary,
it gives a small imaginary part to the two existing poles,
\beq
k_I^{(1)} \sim \frac{M_{lo}^2}{M_{hi}}, 
\eeq
so that $\Gamma \sim M_{lo}^3/(m M_{hi})$
in Eq. (\ref{smat}).
For $l\ge 1$, the last two terms in the denominator
bring higher powers of $k$ and thus, in addition
to a small shift of the real part of the existing poles to
$\pm k_R^{(1)}$, they introduce
new poles $k_i$.
These are deep poles since they are at a relatively
large distance from the origin, $|k_i|\sim M_{hi}$.
They are in the region where the EFT breaks down
and therefore cannot be attributed any concrete physical meaning.
They contribute to the smooth background $s_l(k)$ in Eq. (\ref{smat}). 

Of course, 
the power counting we formulated led to the perturbative expansion
of $T$ when $Q\sim M_{lo}$. 
The resummation we carried out in Eq. (\ref{geo}) was 
in principle unnecessary.
Strictly speaking,
in our perturbative scheme, the pole position does not shift.
Instead, multiple poles at the original
location are generated at higher orders.

However, right on the resonance, the formally higher-order terms
cannot be neglected in Eq. (\ref{fullt}) because
the formally leading-order terms vanish.
In fact, in a momentum region of $O(M_{lo}^{2}/M_{hi})$ around the resonance
the formally leading term is kinematically
suppressed by a factor $\sim (M_{lo}/M_{hi})$.
In this region we have to reorganize the 
expansion, treating $g_l$ as a leading interaction.
In particular, no longer can the unitarity correction induced
by the bubble be treated in perturbation theory.
It is this second, ``kinematic'' fine-tuning that leads
to the resummation in Ref.~\cite{halos}.

Let us illustrate these features in the simplest cases.
For an $s$ wave, 
the poles are the roots of
\beq
-\frac{1}{a_0} + \frac {r_0}{2} k^2 -ik +\ldots=0.
\eeq
The shallow poles have an expansion
\beq
k_\pm =\pm \sqrt{\frac{2}{a_0 r_0}} 
       +\frac{i}{r_0}+\ldots
\eeq
An example of a shallow $s$-wave resonance is elastic $\alpha\alpha$ 
scattering,
but it requires a simultaneous account of the Coulomb interaction.
An EFT approach to this reaction is in progress \cite{gelman}.

For a $p$ wave, the poles come from
\beq
-\frac{1}{a_1} + \frac {r_1}{2} k^2 -ik^3 +\ldots=0.
\eeq
The shallow poles have a similar expansion,
\beq
k_\pm =\sqrt{\frac{2}{a_1 r_1}} 
       \left[\pm 1 +\frac{i}{r_1}\sqrt{\frac{2}{a_1 r_1}} +\ldots \right].
\eeq
Analogous expansions in $M_{lo}/M_{hi}$ can be
written for $E_0$ and $\Gamma(E_0)$.
Note that a third pole appears in the $p$-wave case at 
\bea
&k_1 = \frac{i}{6} |r_1| \left( 1 +\frac{|a_1|^{1/3} |r_1|}{v}
         +\frac{v}{|a_1|^{1/3}|r_1|} \right),&
\nonumber\\
&v=\left(108+|a_1||r_1|^3 +108\sqrt{1+|a_1||r_1|^3/54}\right)^{1/3}.&
\eea
This is a deep ($|k_1|\sim M_{hi}$) bound state.
For higher waves at least a $k^4$ term appears in the pole equation, 
leading to further deep poles and further contributions to the smooth
background in Eq.~(\ref{smat}). 

An example of a shallow $p$-wave resonance comes from 
elastic $n\alpha$ scattering.
Waves that contribute to this process at low energies
are $s_{1/2}$, $p_{1/2}$, $p_{3/2}$, {\it etc.}, which we denote
$0+$, $1-$, $1+$, {\it etc.} as in Ref.~\cite{halos}. 
The phase shift in the $p_{3/2}$ rises quickly from threshold,
and crosses $\pi/2$ near 1 MeV; correspondingly the total
cross section has a bump around this energy. Meanwhile, other
phase shifts evolve more smoothly.
Using our power counting estimates,
$|a_{0+}|\sim |r_{0+}| \sim 1/M_{hi}$,
$|a_{1-}|\sim 1/M_{hi}^3$, $|r_{1-}| \sim M_{hi}$,
$|a_{1+}|\sim 1/M_{hi}M_{lo}^2$, $|r_{1+}| \sim M_{hi}$,
and the values for the scattering parameters in the phase-shift
analysis of Ref.~\cite{ALR73},
we find $M_{hi}\sim 100$ MeV and $M_{lo}\sim 30$ MeV.
The $T$-matrix can be calculated in a way similar to
Ref.~\cite{halos} ---the generalization to distinct fields
and spin is straightforward.
In that reference the resummation was carried out at all
momenta $Q\sim M_{lo}$.
We now compare that with the minimal approach
where the resummation is not done.

The $M_{lo}/M_{hi}$ expansion works reasonably well for
some low-energy quantities.
For example,
the first term in the expansion 
of the energy $E_0$ of the resonance
turns out
to be 0.96 MeV, not very far from the 0.80 MeV found
in next-to-leading order in  Ref.~\cite{halos}.
Likewise, the first term in the expansion
of the width $\Gamma(E_0)$ at the resonance
is 0.82 MeV, to be compared with 0.55 MeV \cite{halos}.

\begin{figure}[tb]
\begin{center}
\includegraphics[width=5in,angle=0,clip=true]{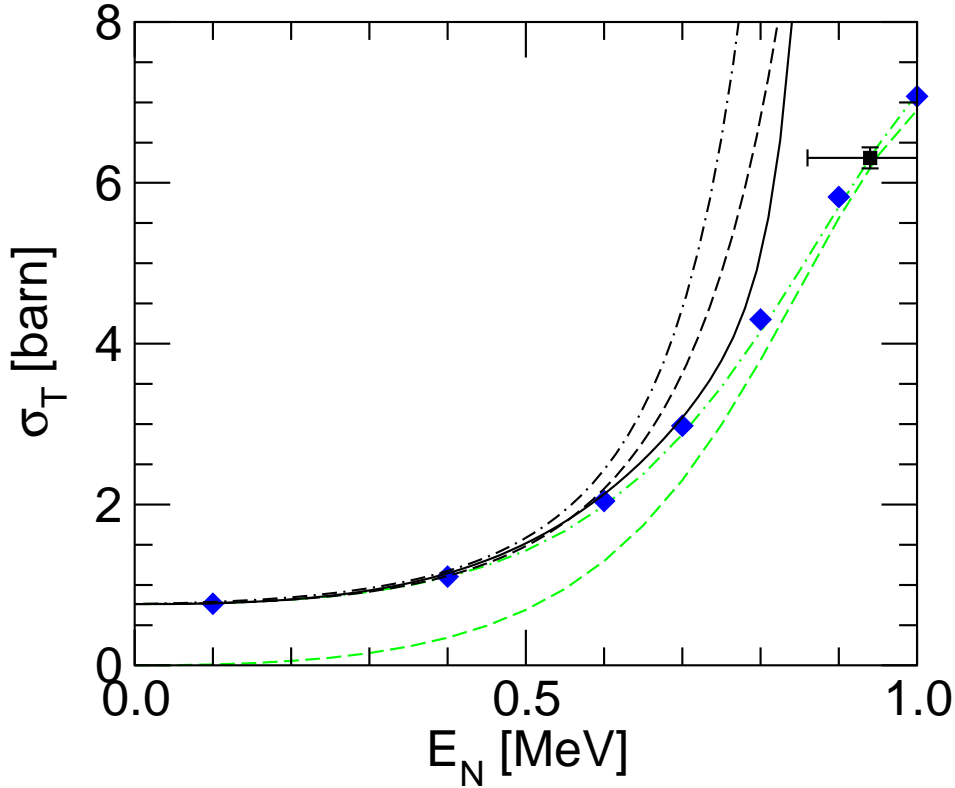}
\end{center}
\caption{The total cross section for $n\alpha$ scattering
 (in barns) as a function of the neutron kinetic energy in the
 $\alpha$ rest frame (in MeV). The diamonds are evaluated data from
 Ref.~\cite{BNL}, and the black squares are experimental 
 data from Ref.~\cite{data}.
 The dashed, dash-dotted, and solid black lines show the result in the 
 EFT without resummation at LO, NLO, and NNLO, respectively. The 
 grey dashed and dash-dotted lines show the result in the EFT with  
 resummation at LO and NLO, respectively.
}
\label{fig:sigtot}
\end{figure}

We show in Fig. \ref{fig:sigtot} the results for the total cross section
as a function of the neutron kinetic energy in the
$\alpha$ rest frame.
The diamonds are ``evaluated data points'' from Ref.~\cite{BNL}.
In order to have an idea of the error bars from individual experiments
we also show data from Ref.~\cite{data} as the black squares.
The EFT in leading (LO), subleading (NLO), and subsubleading (NNLO) orders 
is represented by the dashed, dash-dotted, and solid black lines, 
respectively.   
At LO the scattering length and effective range in the $p_{3/2}$
partial wave as well as the scattering length in the $s_{1/2}$
partial wave contribute. The data are reproduced up to neutron
energies of about $E_N=0.5$ MeV.
Interestingly, the NLO result, which contains only the leading unitarity
correction to the LO result and adds no new parameters, worsens the 
description of the data. At NNLO three more parameters enter:
the shape parameter in the $p_{3/2}$ wave, the effective range in the 
$s_{1/2}$ wave, and the scattering length in the $p_{1/2}$ wave.
The data are described up to $E_N \approx 0.8$ MeV at NNLO.
As expected, the EFT describes the data qualitatively,
but it fails in the immediate neighborhood of the resonance.
In order to improve the description in this neighborhood,
we need to resum the interaction that gives rise to the resonance width.
The calculation then has to be organized in accord to Ref.~\cite{halos}.
For comparison, we show the first two orders of the resummed EFT 
as the grey dashed (LO) and dash-dotted (NLO) lines \cite{halos}.
The improvement around the resonance is evident.
Because the scales $M_{lo}$ and $M_{hi}$ are not very well separated,
that is, $M_{lo}/M_{hi}$ is not very small,
the region of improvement is relatively large.
The resummation is useful throughout the low-energy region. 

\begin{figure}[tb]
\begin{center}
\includegraphics[width=5in,angle=0,clip=true]{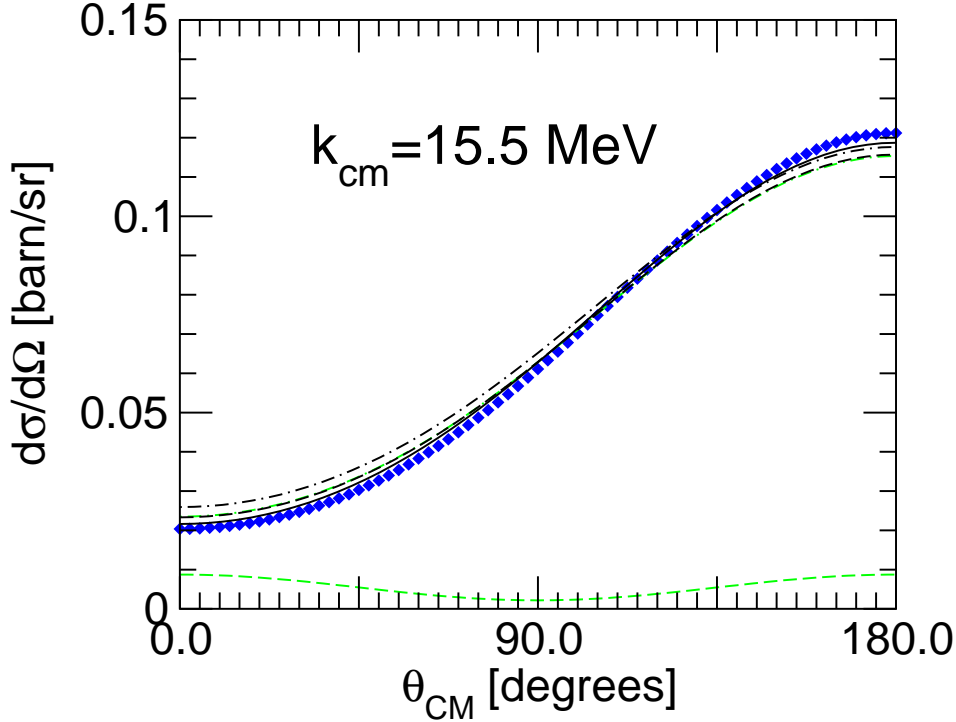}
\end{center}
\caption{The differential cross section for $n\alpha$ scattering
 in the center-of-mass frame (in barns/sr) 
 as a function of the scattering angle $\theta_{cm}$ at a momentum
 of $k_{cm}=15.5$ MeV. The diamonds are evaluated data from
 Ref.~\cite{BNL}.
 The dashed, dash-dotted, and solid black lines show the result in the 
 EFT without resummation at LO, NLO, and NNLO, respectively. The 
 grey dashed and dash-dotted lines show the result in the EFT with  
 resummation at LO and NLO, respectively.
}
\label{fig:sigdiff}
\end{figure}

In Fig. \ref{fig:sigdiff} the results for the differential
cross section in the center-of-mass frame are shown
as a function of the scattering angle $\theta_{cm}$
at a momentum of $k_{cm}=15.5$ MeV (corresponding to
$E_N=0.2$ MeV in the $\alpha$ rest frame).
The diamonds are evaluated data from Ref.~\cite{BNL}.
The EFT in leading (LO), subleading (NLO), and subsubleading (NNLO) orders 
is represented by the dashed, dash-dotted, and solid black lines, 
respectively. The first two orders of the resummed EFT are shown as
the grey dashed (LO) and dash-dotted (NLO) lines \cite{halos}.
Both EFTs describe the differential cross section at 
NLO and higher orders. At LO, however, the resummed EFT badly fails
to reproduce the differential cross section, while the EFT without
resummation already gives a good description. This is because 
in the resummed EFT the $s_{1/2}$ wave is suppressed relative to 
the  $p_{3/2}$ wave and does only enter at NLO. In the EFT without 
resummation, the relative order is changed and both the 
$s_{1/2}$ and $p_{3/2}$  waves enter at LO. 

In summary, we have discussed the treatment of shallow resonances in EFT.
Although the unitarization done in Ref.~\cite{halos}
is not necessary except near the resonance,
it improves the description throughout the low-energy region.
We have considered explicitly only the case
of identical, spinless, heavy particles, but the same ideas
apply to other cases,
such as $\pi N$ scattering near the Delta resonance
---in this context, see Ref.~\cite{dandelta}---
and various low-energy nuclear reactions.
We illustrated this statement by a study of $n\alpha$ scattering.

\section*{Acknowledgments}
UvK is grateful 
to the Nuclear Theory Group at the University of Washington
for its hospitality,
and to RIKEN, Brookhaven National Laboratory and 
the U.S. Department of Energy [DE-AC02-98CH10886] for providing the facilities
essential for the completion of this work.
This research was supported in part
by the Director, Office of Energy Research, 
Office of High-Energy and Nuclear Physics,
by the Office of Basic Energy Sciences, Division of Nuclear Sciences,
of the U.S. Department of Energy under contract DE-AC03-76SF00098 (PFB),
by a DOE Outstanding Junior Investigator Award (UvK),
and by an Alfred P. Sloan Fellowship (UvK).

\end{document}